\documentclass[%
 reprint,
superscriptaddress,
 amsmath,amssymb,
]{revtex4-1}

\usepackage{graphicx}
\usepackage{dcolumn}
\usepackage{bm}
\usepackage{xcolor}
\usepackage{siunitx}

\begin{document}

\title{Generation of Coherent Phonons via a Cavity Enhanced Photonic Lambda Scheme}
	
\author{J. Bourhill}
\affiliation{ARC Centre of Excellence for Engineered Quantum Systems, Department of Physics, University of Western Australia, 35 Stirling Highway, Crawley WA 6009, Australia}
\affiliation{IMT Atlantique and Lab-STICC (UMR 6285), CNRS, Technopole Brest-Iroise, CS 83818, 29238 Brest Cedex 3, France}

\author{N. C. Carvalho}
\affiliation{Applied Physics Department and Photonics Research Center, University of Campinas, Campinas, SP 13051, Brazil}

\author{M. Goryachev}

\affiliation{ARC Centre of Excellence for Engineered Quantum Systems, Department of Physics, University of Western Australia, 35 Stirling Highway, Crawley WA 6009, Australia}

\author{Serge Galliou}
\affiliation{Department of Time and Frequency, FEMTO-ST Institute, ENSMM, 26 Chemin de l{'}\'Epitaphe, 25000, Besan\c con, France}

\author{M.E. Tobar}
\homepage{michael.tobar@uwa.edu.au}
\affiliation{ARC Centre of Excellence for Engineered Quantum Systems, Department of Physics, University of Western Australia, 35 Stirling Highway, Crawley WA 6009, Australia}

\date{\today}

\begin{abstract}
\noindent \textit{We demonstrate the generation of coherent phonons in a quartz Bulk Acoustic Wave (BAW) resonator through the photoelastic properties of the crystal, via the coupling to a microwave cavity enhanced by a photonic lambda scheme. This is achieved by imbedding a single crystal BAW resonator between the post and the adjacent wall of a microwave reentrant cavity resonator. This 3D photonic lumped LC resonator at the same time acts as the electrodes of a BAW phonon resonator, and allows the direct readout of coherent phonons via the linear piezoelectric response of the quartz. A microwave pump, $\omega_p$ is tuned to the cavity resonance $\omega_0$, while a probe frequency, $\omega_{probe}$, is detuned and varied around the red and blue detuned values with respect to the BAW phonon frequency, $\Omega_m$. The pump and probe power dependence of the generated phonons unequivocally determines the process to be electrostrictive, with the phonons produced at the difference frequency between pump and probe, with no back action effects involved. Thus, the phonons are created without threshold and can be considered analogous to a passive Coherent Population Trapped (CPT) maser scheme.}
\end{abstract}

\maketitle


One of the major hurdles to engineer quantum systems for applications such as sensing and scalable quantum computing is decoherence {--} the computational advantages of entanglement are lost if one's quantum state collapses too quickly. One main pathway of decoherence is the energy lost to the environment. Therefore, investigation into quantum hybrid systems that facilitate the transfer of energy from one form to another, an important protocol for quantum infrastructure, commonly look to utilise high quality factor resonators. As far as mechanical systems go, macroscopic single{--}crystal quartz bulk acoustic wave (BAW) resonators have demonstrated the largest $Q\times f$ products experimentally producible \cite{highq,Gory13,galliouScRip2013,Renninger:2018aa,Khareleaav0582}. These crystals are specifically engineered with a convex curvature that traps phonons in the centre of the resonator, drastically reducing contact losses at its peripheries. Given the piezoelectric nature of quartz, coupling to the acoustic modes is straight forward and can be achieved with an RF source and two electrodes placed on either side of the crystal. Typically these electrodes are placed as close as possible to the crystal to achieve high electromechanical coupling, without touching, to preserve high mechanical $Q$-factors. Given their macroscopic size (weighing on the order of grams) and their excellent frequency stability, these devices have been proposed for use in tests of fundamental physics such as Lorentz invariance \cite{lorentz,Lo:2016aa}, quantum gravity \cite{quantgrav}, high frequency gravity wave detectors \cite{gravwave} and the search for scalar dark matter \cite{Arvanitaki2016}.

Only recently has optomechanical coupling to GHz mechanical modes in such crystals been achieved using two counter-propagating lasers \cite{Renninger:2018aa,Khareleaav0582}. The aforementioned work represents a new form of optomechanical system, and successfully interacts with the quartz mechanical modes without the use of piezoelectricity. Whilst piezoelectricity allows strong electromechanical coupling between photons and acoustic phonons, it is extremely valuable to explore coupling between different frequency ranges of these two interacting energy forms in order to improve the versatility and bandwidth of the quartz BAW as a potential quantum hybrid system. Here, we demonstrate coupling between a microwave resonant cavity and high quality factor quartz BAW resonant modes at MHz frequencies. This demonstration was achieved in two ways, firstly through the generation of mechanical sidebands on the microwave carrier when both phonon and photon modes are driven simultaneously, and secondly by exciting the acoustic mode (monitored via piezoelectricity) using two microwave tones, offset by the mechanical frequency. The latter is inherently a nonlinear effect given two GHz frequency tones produce a MHz frequency acoustic excitation, hence excitation via photoelasticity over piezoelectricity, with the later a linear phenomenon.

Phonon masers (or lasers) have been recently realised and have a clear threshold in generation of phonons in analogy to a photonic maser with population inversion \cite{Vahala:2009kl,PhysRevLett.104.083901,PhysRevLett.110.127202,Navarro-Urrios:2015wk}. In contrast a Coherent Population Trapped (CPT) maser operates in a lambda scheme, where two photon tones excite an atomic transition without population inversion through a non-linear process \cite{cpt1,cpt2,cpt3}. Our system is similar to the CPT maser and has no input power threshold for the generation of phonons, allowing transduction of small microwave signals into mechanical frequencies, potentially at the level of a single phonon. We also note that similar techniques have been used to generate coherent phonons, but from optical frequencies \cite{PhysRevB.83.201103,PhysRevLett.99.217405,Khareleaav0582}.


\begin{figure}[t!]
\centering
\includegraphics[width=0.42\textwidth]{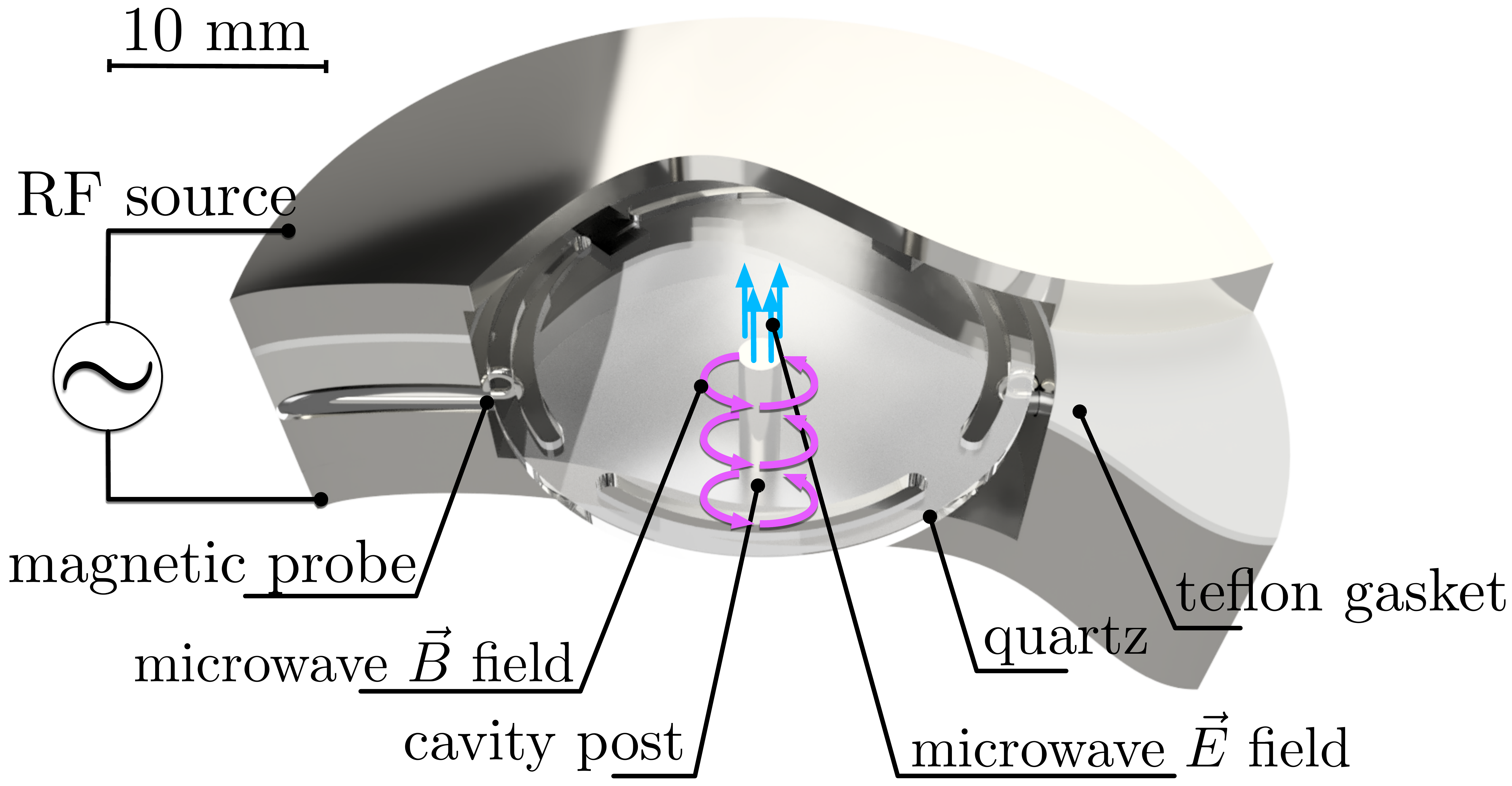}
\caption{Cut-away diagram of the device under test. The teflon gasket electrically isolates the top of the cavity from the bottom, allowing an RF source to piezoelectrically excite mechanical modes in the quartz.}
\label{fig:3D}
\end{figure}

The quartz BAW under study in this experiment was very similar to those used in \cite{highq}: a state of the art SC-cut (Stress Compensated) \cite{quartz1} quartz single crystal utilising BVA technology \cite{quartz2}. The resonator was ultrasonically machined into a planoconvex shape \cite{quartz3}; bulging on one side in its centre. A periphery supporting ring was also machined out of the single quartz crystal, which provides a location to hold the crystal in place. The BAW had a diameter of $d=24.0$ mm, central thickness of $t=1.00$ mm, and radius of curvature of the convex side of $R=300$ mm. Three types of acoustic modes exist; longitudinal modes, fast shear modes, and slow shear modes, or A, B and C-modes, respectively; a result of the anisotropy of quartz. BVA resonators are constructed with {``}non-contacting{''} electrodes placed on either side of the crystal, allowing efficient electromechanical coupling through the quartz{'}s piezoelectricity. These electrodes will only detect a voltage difference across the crystal for an odd harmonic of the A, B or C modes; requiring opposite signed signals at either end of the crystal. Similarly, only odd harmonics can be excited by applying an RF voltage difference to the electrodes. The $3^\text{rd}$ harmonic of the A, B and C modes for the given crystal at 4 K are located at 9.415 143 MHz, 5.500 049 MHz, and 4.996 171 MHz, respectively. The quality factors of these modes improve under vacuum and cryogenic conditions, capable of approaching $\sim10^{10}$\cite{galliouScRip2013}. 

The conducting surfaces of the electrodes potentially interfere with any microwave modes in a cavity QED-like experiment, so the mechanical resonator investigated here does not have any electrodes included. Instead, the top half and bottom half of the microwave cavity were insulated from each other by a teflon layer, allowing electromechanical coupling to the quartz crystal across the two halves of the cavity, hence the cavity can also be utilised as a set of electrodes (see Fig.\ref{fig:3D}). 

The microwave cavity takes the form of a re-entrant, or Klystron cavity \cite{doi:10.1063/1.4848935}: an empty cylindrical space with a conducting post in the centre, which extends from one end-face towards the other, stopping short so as to form a gap between the top of the post and the lid of the cavity. The resonant microwave frequency re-entrant mode is characterised by majority of electric field confined in this gap and the magnetic field circling around the post as shown in Fig.\ref{fig:3D}, and thus the metallic rod forms a 3D lumped element LC resonator. The re-entrant cavity had a resonant frequency of $4.095$ GHz. Microwaves were coupled into and out of the re-entrant cavity via co-axial cables, which were terminated by loops inside the cavity, hence producing an oscillating electromagnetic field. One of the main loss mechanisms at cryogenic temperatures is the surface resistance of the cavity walls. To minimise this, the cavity is constructed from pure Niobium, which becomes superconducting at $\sim9$ K. However, the teflon gasket separating the two halves of the cavity resulted in some losses via leakage, limiting the microwave resonance{'}s $Q$-factor to about $\sim2000$ at 4 K.

The quartz BAW resonator is placed within the gap between the re-entrant cavity{'}s post and lid {--} such that its centre overlaps with the location of highest microwave electric field concentration. This is because the centre of the crystal is the location of its mechanical modes, and a high degree of overlap between the mechanical and microwave modes will result in a larger photoelastic coupling. Recently a similar type of 3D cavity structure was used for transduction from microwave to optical frequencies\cite{Davis2020}. The crystal was supported via three rigid blocks attached to the inside wall of the cavity upon which the peripheral support ring of the BAW makes contact. Due to the photoelastic effect, mechanical strain of the quartz results in a periodic modulation of the dielectric permittivity over the crystal volume. This modulation changes the nature of the media through which the resonant re-entrant mode{'}s electric field is oscillating. This results in a frequency shift of the microwave mode and hence a form of optomechanical coupling. Via the inverse process, electrostriction (or photoelastic response), an applied electric field induces strain within the crystal due to a slight displacement of ions.


\begin{figure}[b!]
\centering
\includegraphics[width=8.5cm]{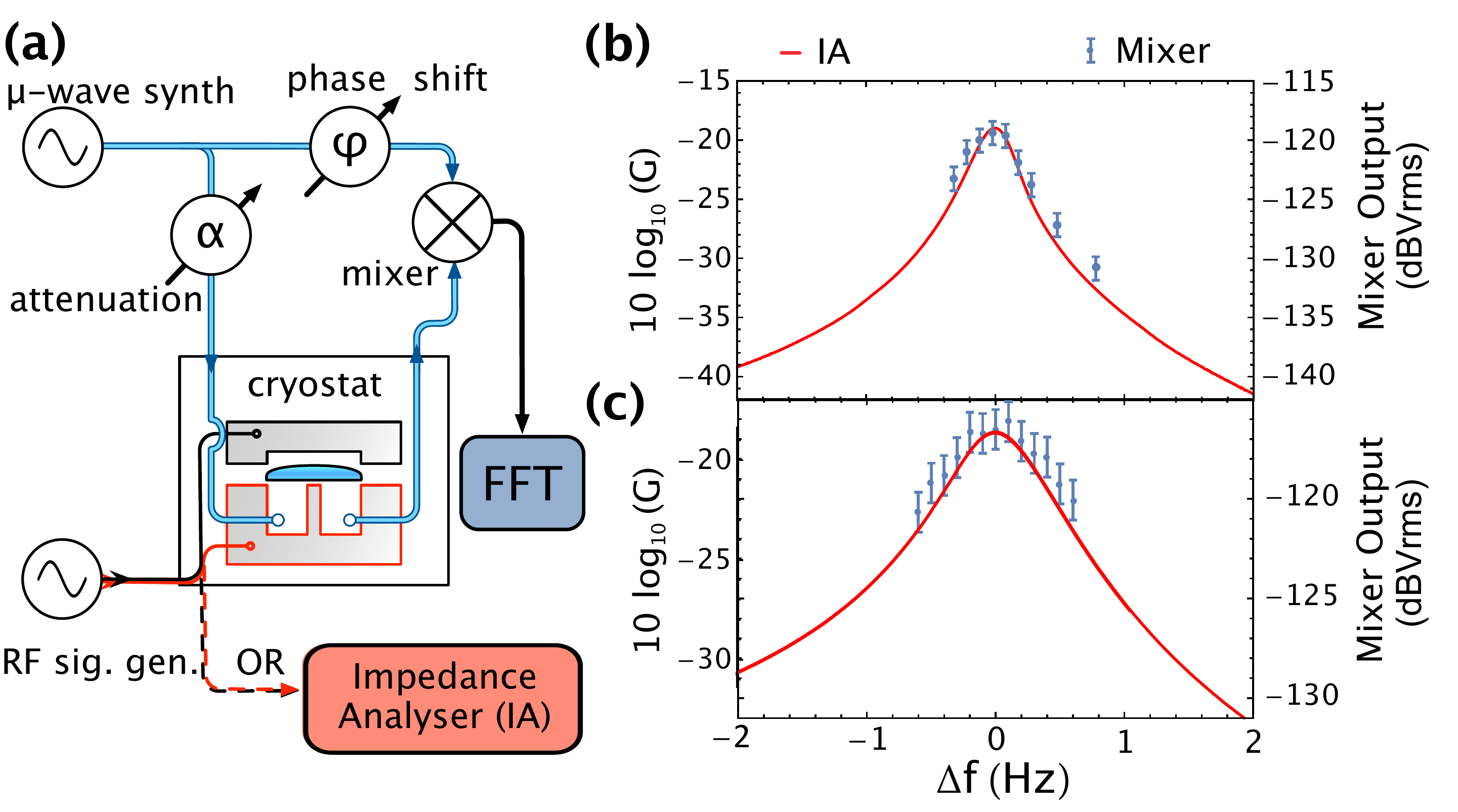}
\caption{(a) Simplified experimental phase bridge setup for observing mechanical sidebands on the microwave carrier and calibrating the electromechanical coupling. (b) ((c)) Mixer output compared to impedance analyser measurement of 4.996 MHz (9.415 MHz) mode. Error bars on the mixer readings are determined from repeat measurements for a given detuning. }
\label{fig:PB}
\end{figure}

            \begin{figure*}
            \centering
            \includegraphics[width=17cm]{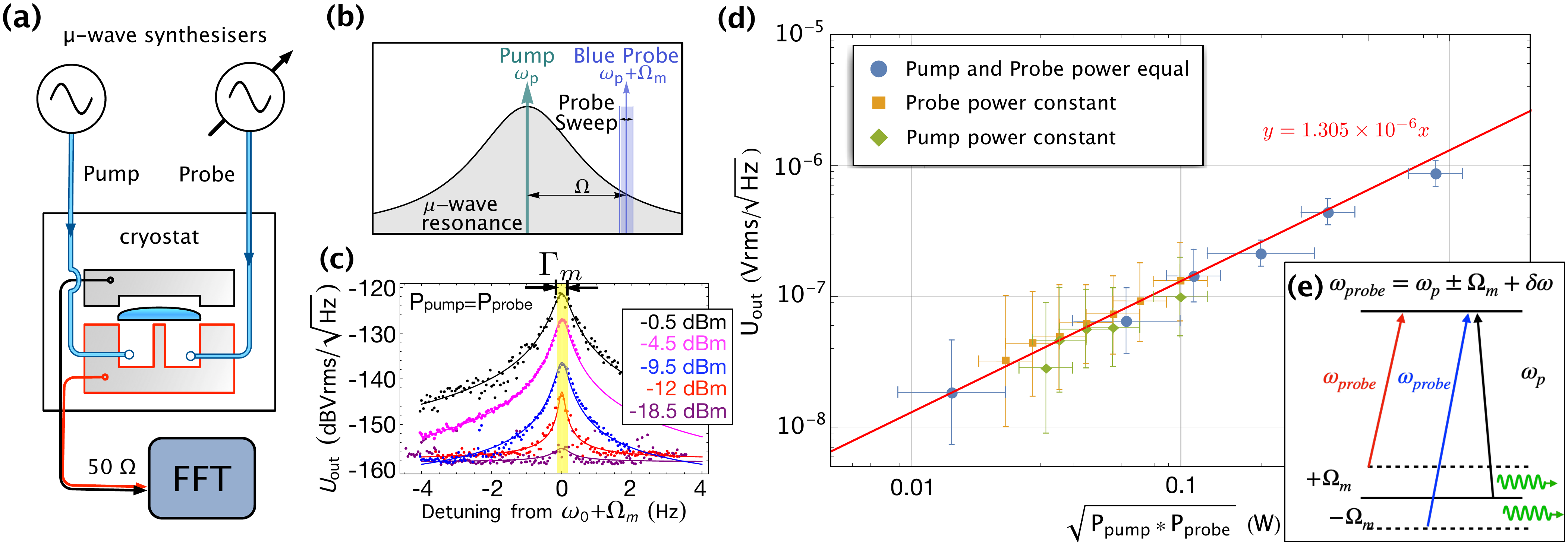}
            \caption{(a) Simplified {``}Lambda-scheme{''} or {``}two-tone{''} experimental setup to generate coherent phonons at the difference frequency of the pump ($\omega_p$) and probe frequency ($\omega_{probe}$). (b) Spectral representation of {``}two-tone{''} experiment for the blue-detuned case. (c) Electrode output as a function of probe detuning from $\left(\omega_p+\Omega_m\right)/2\pi$ for different pump and probe powers ($P_\text{pump}=P_\text{probe}$, values given in the legend) for the $\Omega_m=4.996$ MHz acoustic mode. Each point represents a single experiment that either generates a coherent phonon, or if too weak, measures the back ground noise floor. The reach of the non-linearity can be characterised by a bandwidth, $\Gamma_{excited}$, which depends on power and like most non-linear processes, is wider than the intrinsic line with of the phonon resonance, $\Gamma_{m}$. (d) Electrode output as a function of $\sqrt{P_\text{pump}\times P_\text{probe}}$ when $\delta\omega=0$ (or zero detuning), demonstrating a linear relationship that intercepts $\{0,0\}$, suggesting no threshold power in the process. Horizontal errors are determined from the uncertainty in input and output line attenuations in the cryostat and vertical uncertainties from repeat measurements for given input powers. The equivalent energy diagram similar to a CPT maser is shown in (e).}
            \label{fig:2tone}
            \end{figure*}

The optomechanical coupling was determined by exciting the quartz mechanical modes piezoelectrically, and measuring the effect on the resonant microwave mode using a {``}phase-bridge{''} setup. To do this, a setup like that in figure \ref{fig:PB} (a) was used.  A microwave synthesiser continuously pumps the re-entrant cavity mode at $\sim$4.095 GHz, whilst an RF signal generator applies a MHz signal across the cavity lid and base. The transmitted microwave signal is then mixed down against the input signal from the synthesiser, which is phase shifted such that the mixer will output a voltage proportional to any phase shift produced in the resonator, which is manifested as a frequency shift in the resonator's arm of the bridge. By observing the output spectra of the mixer on an Agilent 89410A Vector Signal Analyser (FFT), one can measure the strength of modulation on the resonant microwave mode caused by the mechanical motion of the quartz. By applying a continuous wave RF voltage across the microwave cavity this can be done in the static regime {-} measuring the mixer output at and around mechanical resonant frequencies. The results of this experiment, sweeping the signal generator over the 4.996 MHz and 9.415 MHz modes are shown in Fig. \ref{fig:PB}(b) and (c), respectively by the blue points. These measurements were taken with synthesiser power at 15 dBm, attenuation to the cavity $\alpha=18$ dBm, and a 10 mV amplitude signal applied by the RF signal generator around the mechanical resonance frequency. The conversion efficiency of the phase-bridge setup (the voltage output by the mixer given some frequency shift in the resonator arm) is measured using an artificial modulation signal to be $d u/d f=11.7$ $\mu$V/kHz. 

The acoustic resonances of the quartz were also directly measured by an impedance analyser connected across the microwave cavity. From the measurements of impedance and phase, one can determine the conductance, $G$ across the {``}electrodes{''}, which is plotted in red for the 4.996 MHz and 9.415 MHz modes in Fig.\ref{fig:PB}(b) and (c), respectively. We see that the static measurements of the phase bridge mixer output match the measurements of $G$ within experimental error, given some coefficient of mixer conversion efficiency. This demonstrates that the modulation of the microwave resonant mode measured by the phase bridge was a result of the mechanical excitation. These measurements also allow an accurate way to determine $Q$ factors; $1.607\times10^7$ for the 4.996 MHz mode and $1.264\times10^7$ for the 9.415 MHz mode, and the $L$, $C$ and $R$ values for the equivalent electrical circuit of the mechanical resonance. These measurements also allowed the determination of the displacement of the quartz crystal when the aforementioned 10 mV signal was applied, and hence the single-photon optomechanical coupling rate, $g_0$. The value $g_0$ represents the frequency shift of the electromagnetic mode caused by the displacement of the mechanical system when a single photon enters the electromagnetic system.

The charge $q$ and displacement $x$ in a piezoelectrical system are related in a linear fashion by an electromechanical coupling constant $k$ \cite{gravwave}:
	\begin{equation}
	q=k x,
	\end{equation}
where 
	\begin{equation}
	k^2=\frac{\Omega_m M_\text{eff}}{Q~R}.
	\end{equation}
Here, $M_\text{eff}$ is the effective mass of the resonance, $\Omega_m/2\pi$ the resonant frequency, and $R$ its effective resistance. For $\Omega_m/2\pi=4.996$ MHz, an rms charge is determined from the applied $V_\text{rms}=10/\sqrt{2}$ mV and the equivalent resistance $R=78.5~\Omega$. $M_\text{eff}=1.13\times10^{-5}$ kg is determined from finite element modelling \cite{natalia}. This nets an rms displacement of $x=5.52$ nm. The simple relationship 
	\begin{equation}
	\frac{\delta u}{\delta x}=\frac{d u}{d f}\frac{d f}{d x}
	\end{equation}
will allow us to relate the calculated displacement $\delta x=x$ to the output voltage of the mixer $\delta u$ from the aforementioned frequency sensitivity of the phase bridge $d u/d f$ and the dependence of the electromagnetic frequency on displacement $d f/d x=g_0/x_\text{zpf}$, where $x_\text{zpf}=\sqrt{2/\hbar\Omega_m M_\text{eff}}$ is the so-called zero-point fluctuation of the mechanical resonance. Substituting in all relevant numerical values nets a single-photon optomechanical coupling rate of $g_0=8.43$ nHz, which is in excellent agreement with simulated results of the photoeleastic coupling rate in this system \cite{natalia}.

       
The second technique used to excite the mechanical modes involved two microwave tones in a Brillouin-like setup \cite{Renninger:2018aa}. The two signals were input to the microwave cavity via loop probes, and the spectra of the potential difference across the cavity was measured on a FFT spectrum analyser as shown in figure \ref{fig:2tone}(a). The output voltage measured in this way was therefore directly proportional to the displacement of the piezoelectric quartz. 

The FFT window was centred at the acoustic resonant frequency with a 10 Hz span; aiming to detect the voltage spectra produced by the mechanical motion of the piezoelectric quartz. One microwave signal; the pump, was tuned on resonance $\omega_p=\omega_0$, whilst the other; the probe, was detuned by some amount. The probe microwave source was swept from $f_1=\left(\omega_p-\Omega_m\right)/2\pi-5$ Hz to $f_2=\left(\omega_p-\Omega_m\right)/2\pi+5$ Hz in the red detuned case, and $f_1=\left(\omega_p+\Omega_m\right)/2\pi-5$ Hz to $f_2=\left(\omega_p+\Omega_m\right)/2\pi+5$ Hz in the blue detuned case. This scheme is demonstrated in figure \ref{fig:2tone}(b) for the blue-detuned case. Results of the red-detuned case are identical. 

\begin{figure}[b!]
\includegraphics[width=8.5cm]{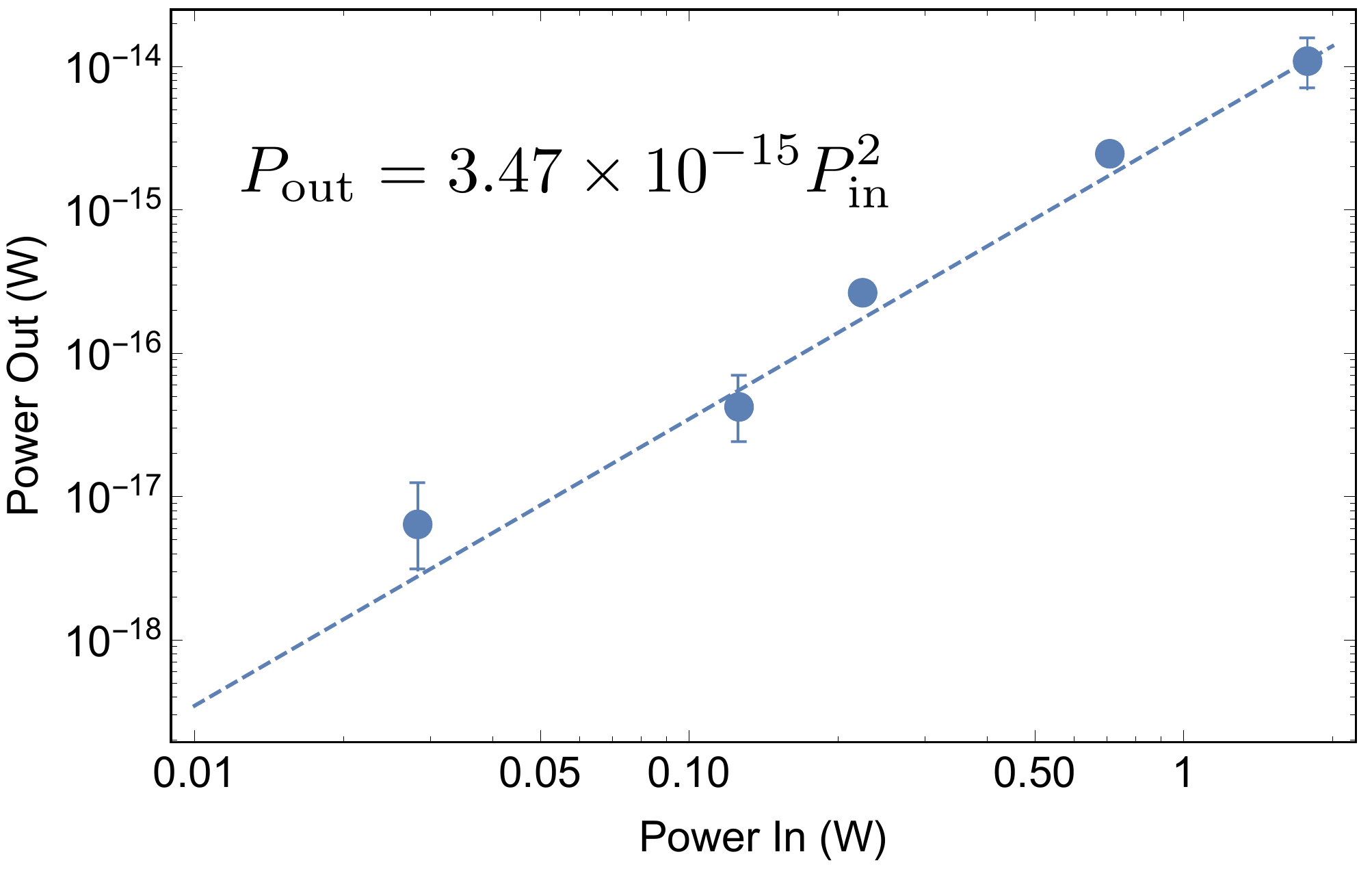}
\caption{Output power at $\Omega_m$ as measured on the FFT device as a function of incident power at $\omega_p$.}
\label{fig:power}
\end{figure}

The output voltage produced by the quartz crystal is plotted in fig \ref{fig:2tone}(c) as the probe synthesiser is detuned from $\omega_0+\Omega_m$. Mechanical motion of the quartz is generated through nonlinear mixing the two microwave input signals, a result of electrostriction. Electrostriction (or photoelasticity) is a quadratic phenomenon that relates strain to the square of electric polarisation according to:
\begin{equation}
S_{ij}=Q_{ijkl}\chi^2\epsilon_0^2E_kE_l,
\label{eq:es}
\end{equation}
where $S_{ij}$ is the second-order strain tensor, $Q_{ijkl}$ the four rank electrostriction coefficient, $\chi$ the electrical susceptibility (can be simplified to a scalar) and $E_k$, $E_l$ electric fields. 
The quadratic nature of the electrostriction generates a double frequency and difference frequency term. When the difference frequency is equal to a mechanical resonant frequency of the quartz crystal, it will be resonantly enhanced. Given the piezoelectric nature of the quartz, this will generate an electric field at the same difference frequency, and hence a voltage across the {``}electrodes{''}; i.e. the top and the bottom of the cavity.

The strain produced by electrostriction acts as a driving term in the piezolectric equations of motion \cite{tiersten}. A full theoretical derivation of this process is given in the supplementary materials, which demonstrates that the voltage across the top and bottom of the cavity is $U_\text{out}\propto E_\text{pump}E_\text{probe}\propto\sqrt{P_\text{pump}P_\text{probe}}$. The dependence of $U_\text{out}$ on the pump and probe powers is plotted in figure \ref{fig:2tone}(d), demonstrating this proportionality. The nonlinear process described here necessitates that the coherence of the pump and probe signals is maintained by the generated phonons. 

From the above relationship, we can expect a quadratic dependence of output power on input power as demonstrated by figure \ref{fig:power}. Output power is derived from the measurements in figure \ref{fig:2tone}(c). The efficiency is very low given the small value of $g_0$; a result of lower optical and mechanical frequencies relative to previous publications \cite{PhysRevB.83.201103,PhysRevLett.99.217405,Khareleaav0582}, and suboptimal electromechanical coupling to the quartz piezoelectric current, resulting in some of the signal being lost at the readout stage.

Nevertheless, the lack of apparent threshold for the generation of phonons means that the smallest possible detected signal was determined by the noise floors, which include the instrument readout, the noise temperature of amplification.

\begin{figure}[t]
\includegraphics[width=8.5cm]{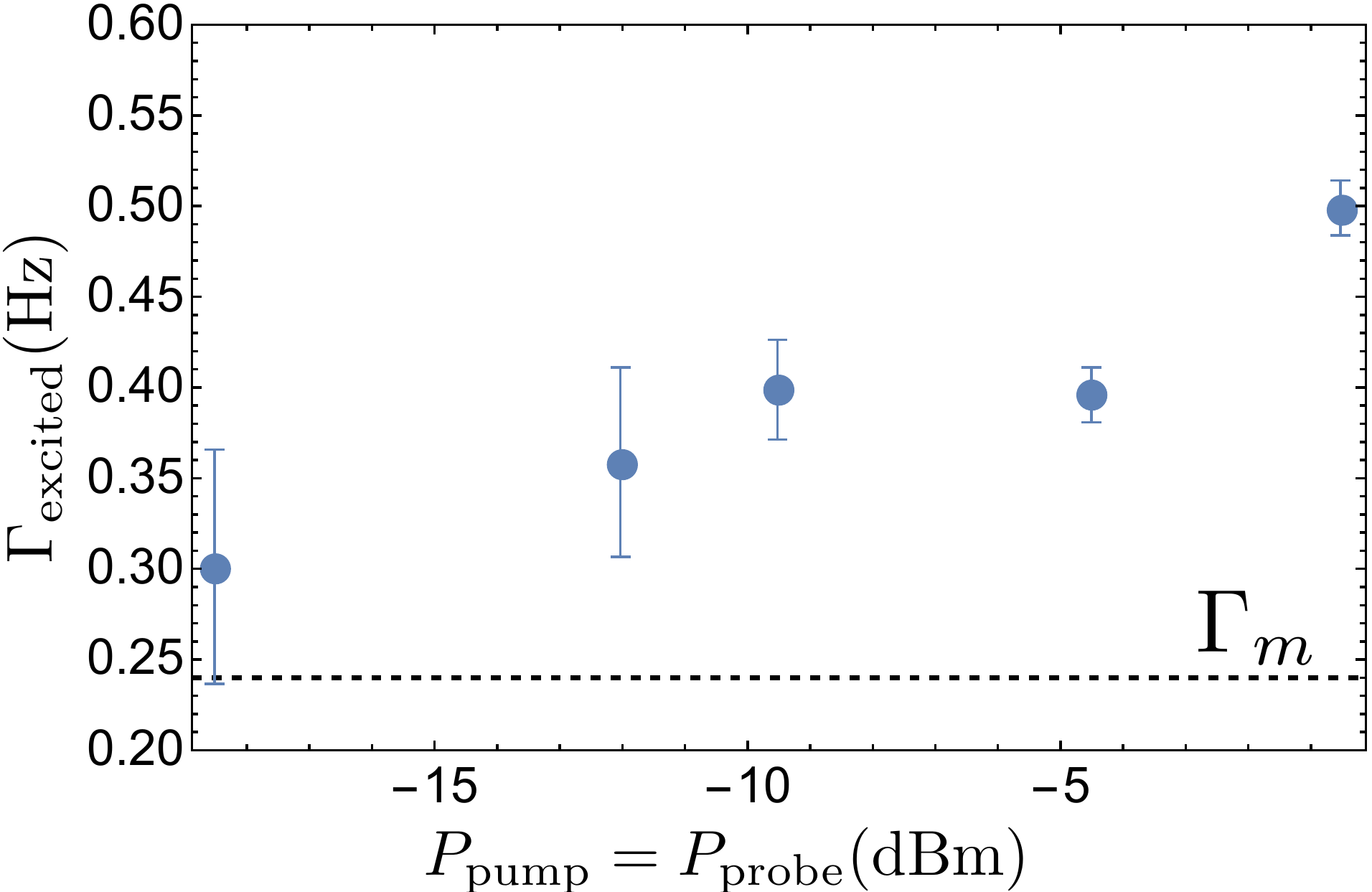}
\caption{3dB bandwidth of the frequency range of the phonon excitation process $\Gamma_\text{excited}$ as a function of applied microwave power (also see Fig.\ref{fig:2tone}(c)). The frequency range of the possible phonon generation is centred around the BAW acoustic frequency and is larger than the intrinsic bandwidth $\Gamma_m=\Omega_m/Q_m$ of the BAW acoustic mode. The frequency range increases as a function of power and for the range of powers we applied it was more than a factor of two greater than the intrinsic linewidth. Error bars are estimated from fitting Lorentzian functions to the frequency profile of the phonon excitations in Fig.\ref{fig:2tone}(c). }
\label{fig:bw}
\end{figure}

Given the microwave mode is a lumped resonator, the typical phase matching conditions of a Brillouin scheme are lifted. Instead, all that is necessary for the generation of mechanical phonons is conservation of energy between the two microwave fields and the acoustic mode, with the generated phonons directly detected through the direct electrical readout of the piezoelectric quartz. This scheme is analogous to a passive CPT maser \cite{cpt1,cpt2,cpt3}, in which two detuned optical pumps generate a microwave signal near the frequency of a hyperfine splitting, but determined by the frequency difference of the optical pumps due to the non-linearity.  Exactly like CPT maser, the observed excitation demonstrates no threshold, as shown by Fig.\ref{fig:2tone}(d). The allowable frequency range of the excited coherent phonons (Fig.\ref{fig:2tone}(c)) increases as a function of applied power, demonstrated in Fig.\ref{fig:bw}.

In conclusion we have demonstrated a way to calibrate an optomechanical system coupled through a non-linear electrostrictive coupling. By implementing a two-photon lambda excitation coherent phonons were generated, with a bandwidth of generation greater than the acoustic linewidth. This technique is analogous to a CPT maser, and gives a way to generate coherent phonons with no power threshold.

The data that support the findings of this study are available from the corresponding author upon reasonable request.

\section*{Supplementary Material}
Please find a complete theoretical derivation of the photoelastic coupling mechanism described above in the Supplementary Material. Here, it is shown how the dependence of the output voltage measured of the quartz resonator in the two tone experiment is proportional to the square root of the product of the two incident powers of the microwave signals. \\
\\
This work was supported by Australian Research Council grant number CE170100009 and a UWA Research Collaboration Award with IMT-Atlantique.

\newcommand{\beginsupplement}{%
        \setcounter{table}{0}
        \renewcommand{\thetable}{S\arabic{table}}%
          \renewcommand{\theequation}{S\arabic{equation}}%
        \setcounter{figure}{0}
        \renewcommand{\thefigure}{S\arabic{figure}}%
     }

 \newpage
 ~
 \newpage
\beginsupplement
 
\section*{Supplementary Material: Output voltage from two applied microwave sources}

We drive the microwave cavity with two tones, $E^\text{pump}$ at frequency $\omega_0$ and $E^\text{probe}$ at frequency $\omega_0+\Delta$, where $\omega_0$ is the microwave resonant frequency of the cavity. Given the re-entrant cavity architecture, the electric field exists between the top of the post and the roof of the cavity and can therefore be approximated as the electric field inside a capacitor; i.e. in the cavity{'}s $z$-direction and uniform throughout. Therefore we can write the electric fields between the top of the post and the roof of the cavity as:
\begin{multline}
E^\text{pump}(t)=E_0^{(1)}(e^{i\omega_0t}+e^{-i\omega_0t}).\bf{\hat{e}_z}~\text{and}~\\E^\text{probe}(t)=E_0^{(2)}(e^{i(\omega_0+\Delta)t}+e^{-i(\omega_0+\Delta)t}).\bf{\hat{e}_z},
\end{multline}
where $E_0^i$ is the electric field amplitude and $\phi$ and $\theta$ represent the phase of both signals. 

%

The two electric fields across the quartz crystal will interact via electrostriction; a quadratic phenomenon that relates strain to the square of electric polarisation P, according to:
\begin{equation}
S_{ij}=Q_{ijkl}\text{P}_k\text{P}_l=Q_{ijkl}\chi^2\epsilon_0^2E_kE_l=M_{ijkl}E_kE_l,
\label{eq:es}
\end{equation}
where $S_{ij}$ is the second-order strain tensor, $Q_{ijkl}$ the four rank electrostriction tensor coefficient, $\chi=\epsilon_r-1$ the electrical susceptibility and $E_k$, $E_l$ electric fields. Substituting $E^{(1)}$ and $E^{(2)}$ into \ref{eq:es} gives:
\begin{multline}
S_{ij}=Q_{ijkl}\chi^2\epsilon_0^2 E_0^{(1)}E_0^{(2)}(e^{i(2\omega_0 +\Delta)t}+e^{i\Delta t}+c.c.).\bf{\hat{e}_z}.\bf{\hat{e}_z}\\
=K_{ijkl}E_0^{(1)}E_0^{(2)}[(e^{i(2\omega_0 +\Delta)t}+e^{i\Delta t}+c.c.)].\bf{\hat{e}_z}.\bf{\hat{e}_z}
\label{eq:5}
\end{multline}

\begin{figure}[t!]
\includegraphics[width=\columnwidth]{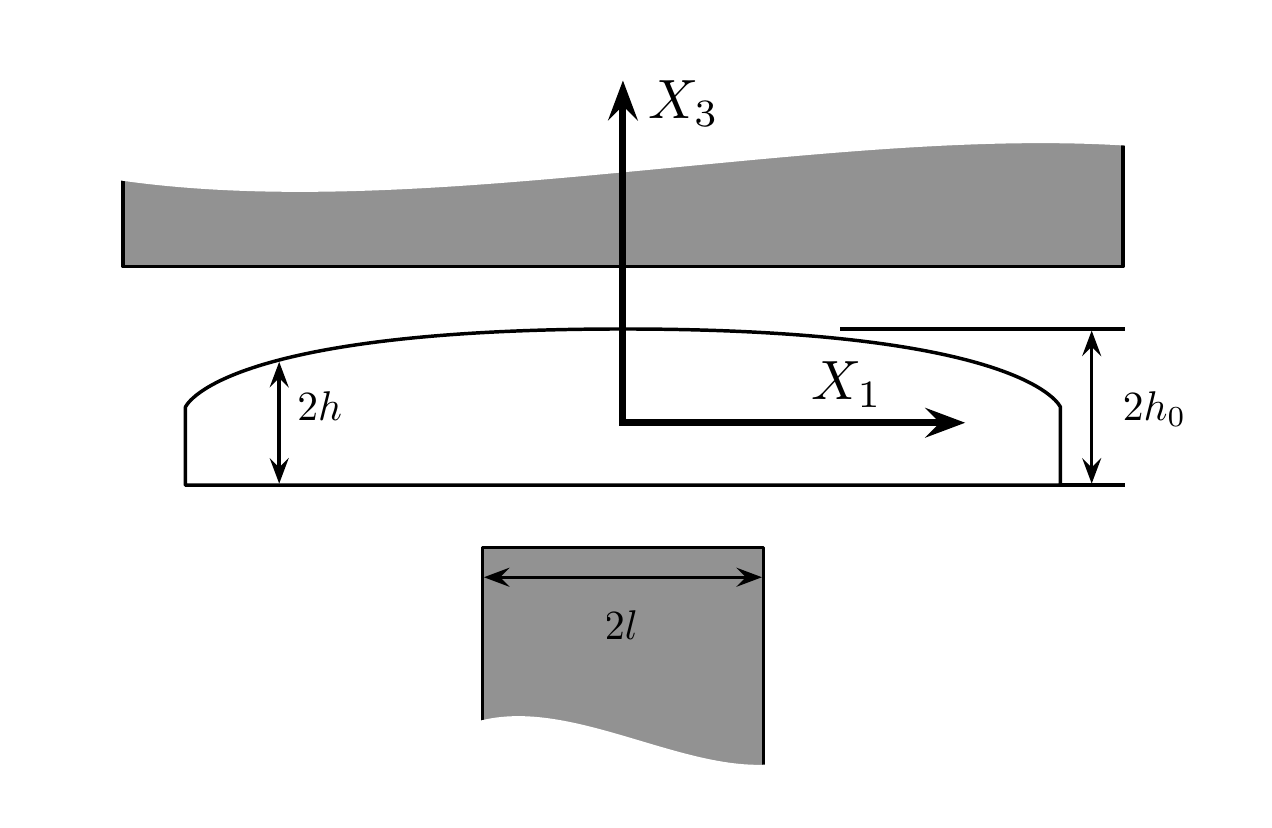}
\caption{Schematic diagram showing a cross section of the plano-convex quartz resonator inside the re-entrant cavity}
\label{fig:diag}
\end{figure}

Given the two applied $E$-fields are forced into the $z$-direction ($X_3$ in figure \ref{fig:diag}) by the cavity, it follows that strain components $S_{ij}$ will be determined by the electrostriction tensor components $Q_{ij33}$ (modified in this case to $K_{ij33}$). If $\Delta=\Omega_m^i$ (i.e. $\Delta<<\omega_0$), the frequency of a resonant mechanical BAW mode in the quartz crystal, the low frequency component in equation \ref{eq:5} will generate strain at the frequency $\Omega_m^i$ which will be resonantly enhanced, and hence we can ignore the higher frequency term, which will not be seen by the mechanical system. This strain term will appear as an additional term in the standard piezoelectric differential equation of motion \cite{tiersten}:
\begin{multline}
T_{ij}=c_{ijkl}\frac{\partial u_{k}}{\partial x_l}+e_{kij}\frac{\partial \varphi}{\partial x_{k}},\\
D_i=e_{ikl}\frac{\partial u_{k}}{\partial x_l}-\epsilon_{ij}\frac{\partial\varphi}{\partial x_{j}},\\
\frac{\partial T_{ij}}{\partial x_i}=\rho\ddot{u}_j,~\frac{\partial D_i}{\partial x_i}=0,
\label{eq:piezo}
\end{multline}
where $T_{ij}$, $u_j$, and $D_i$ are the components of stress, mechanical displacement, and electric displacement, respectively; $\rho$ and $\varphi$ are the mass density and electric potential, respectively; and $c_{ijkl}$, $e_{ikl}$ and $\epsilon_{ij}$ are the elastic, piezoelectric, and dielectric constants, respectively.

Given we deal with a plano-convex crystal architecture (see fig. \ref{fig:diag}) and are interested in only the third overtone of the pure shear mode (with resonant frequency $\Omega_m^{(300)}$) the equation of motion for displacement becomes \cite{tiersten} (setting the tunable $\Delta=\omega$):
\begin{multline}
M_3\frac{\partial^2u_1}{\partial x_1^2}+P_3\frac{\partial^2u_1}{\partial x_2^2}-\frac{3^2\pi^2\bar{c}^{(1)}}{4h_0^2}\left(1+\frac{x_1^2+x_2^2}{2 R h_0}\right)u_1\\+\rho\omega^2 u_1=-\rho\omega^2K_{1133}E_0^{(1)}E_0^{(2)}e^{i\omega t},
\label{eq:eom}
\end{multline}

where $u_1=u(x_1,x_2)e^{i\omega t}$ is the component of mechanical displacement in the $X_1$ direction; $M_n$, $Q_n$ and $P_n$ are constants formulated from various elastic tensor and dimensional values \cite{tiersten}; $h_0$ and $R$ are the height of the crystal at its centre and its radius of curvature, respectively; and $\bar{c}^{(1)}=\rho \omega^2/\eta_1^2$, where $\eta_1=\omega/v_1$, where $v_1$ is the speed of the acoustic wave in the $X_1$ direction. 


Equation \ref{eq:eom} will have solutions of the form:
\begin{multline}
u_1=H^{300}u_{300}\sin\frac{3\pi x_3}{2h}e^{i\omega t},\\
\varphi=\frac{e_{36}}{\epsilon_{33}}H^{300}u_{300}\left(\sin\frac{3\pi x_3}{2h}+\frac{x_3}{h}\right)e^{i\omega t},\\
\text{where}~H^{300}=\frac{-K_{1133}E_0^{(1)}E_0^{(2)}\sqrt{\alpha_3}\sqrt{\beta_3}4F_{130}F_{330}}{1-\Omega_m^2/\omega^2},\\
F_{130}=\int_0^{l_1}e^{-\alpha_3(x_1^2/2)}dx_1,~F_{330}=\int_0^{l_2}e^{-\beta_3(x_2^2/2)}dx_2\\
u_{300}=e^{-\alpha_3(x_1^2/2)}e^{-\beta_3(x_2^2/2)},\\
\alpha_n^2=\frac{n^2\pi^2\bar{c}^{(1)}}{8Rh_0^3M_n},~\beta_n^2=\frac{n^2\pi^2\bar{c}^{(1)}}{8Rh_0^3P_n},
\end{multline}
and, as usual, $\Omega_m$ is replaced by
\begin{equation}
\hat{\Omega}_m=\Omega_m+i\Omega_m/2Q,
\end{equation}
in which $Q$ is the unloaded quality factor of the resonant mode. 
From equation \ref{eq:piezo}, and assuming some scalar potential $\varphi$ exists from the resulting electric field where quantities are varying at $\omega$, and various simplifications \cite{tiersten} we can state that
\begin{equation}
D_3=e_{36}\frac{\partial u_1}{\partial x_3}-\epsilon_{33}\frac{\partial \varphi}{\partial x_3}
=\frac{-e_{36}H^{300}u_{300}}{h_0}e^{i\omega t},
\end{equation}

from which the current generated by the crystal{'}s motion can be obtained by integrating over the crystal area:

\begin{equation}
I=-\int_{A_e}\dot{D}_3dx_1dx_2
=\frac{i\omega e_{36} H^{300}}{h_0}\int_{A_e}u_{300}dx_1dx_2,
\label{eq:current}
\end{equation}
where $A_e$ is the area of the {``}electrodes{''}, and we are able to remove all terms independent of $x_1$ and $x_2$ from the integration, which assumes the electric fields $E_0^i$ are constant over the area of the crystal.

The current generated by the crystal between the electrodes is received at the 50$\Omega$ terminal of the FFT device as a voltage $U_\text{out}$. We can see from the from of equation \ref{eq:current} that the amplitude of this voltage will be proportional to $H_{300}$ and hence $E_0^{(1)}E_0^{(2)}$ when $\omega=\Omega_m$ i.e. the detuning $\Delta=\Omega_m$.

Given we are assuming these fields are approximately those of a parallel plate capacitor we can relate the amplitude of the fields to the input power of the driving signals by considering the entire microwave cavity system as a lumped parallel LCR circuit and therefore the voltage drop across the capacitor to its reactance, $X_c(\omega)$ and the input powers, $P^i$ :
\begin{equation}
E_0^{(1)}=\frac{\sqrt{X_c(\omega_0)P^{(1)}}}{d},~\text{and}~E_0^{(2)}=\frac{\sqrt{X_c(\omega_0+\Delta)P^{(2)}}}{d},
\end{equation}
where $d$ is the distance between the top of the post and the roof.

Therefore, we can state that the voltage $U_\text{out}$ read from the FFT will have the proportionality:
\begin{equation}
U_\text{out}\propto \sqrt{P^{(1)}P^{(2)}}
\end{equation}

\end{document}